\documentclass{ws-p9-75x6-50}

\begin{document}

\title{Supermassive boson stars: prospects for observational detection}

\author{Diego F. Torres}

\address{Instituto Argentino de Radioastronom\'{\i}a \\C.C.5,
1894 Villa Elisa, Buenos Aires, Argentina\\ E-mail:
dtorres@venus.fisica.unlp.edu.ar}

\author{S. Capozziello and G. Lambiase}

\address{Dipartimento di Scienze Fisiche E.R.
Caianiello, Universit\`a di Salerno\\ 84081 Baronissi (SA),
Italy\\ \& \\Istituto Nazionale di Fisica Nucleare, Sez. Napoli,
Italy
\\Email: capozziello@sa.infn.it; lambiase@sa.infn.it}


\maketitle

\abstracts{In a recent work (Torres, Capozziello and Lambiase,
Physical Review D62, 104012 (2000)), it was shown that a
supermassive boson star could provide an alternative model for the
galactic center, usually assumed as a black hole. Here we comment
on some of the possibilities to actually detect this object, and
how can it be distinguished from the standard and other
alternative models.}

Although it is commonly believed that the center of the Galaxy is
a supermassive black hole, it is not yet established on a firm
observational basis. We have recently shown \cite{T} that a
supermassive boson or soliton star, if it exists at all, could
well be at the center of some galaxies, including ours. These
models can be in agreement current observations, and fit very well
into dynamical and gravitational requirements. Their main features
are:
\begin{enumerate}
  \item The central object is highly relativistic, with a size comparable to (but
  slightly larger than) the Schwarzschild radius of a black hole
  of equal mass.
  \item It has neither an event horizon nor a singularity,
  and after a physical radius is reached, the
  mass distribution exponentially decreases.
  \item The particles that form the object interact between each other
  only gravitationally, in such a way that there is no solid
  surface to which falling particles can collide.
\end{enumerate}
Then, how can we differentiate between this and the standard
model? Without entering into the details, referring the interested
reader to Ref. \cite{T} and references therein, we shall mention
some of the possibilities.

Firstly, one can make an in-depth {\it study of the properties of
the accretion disk}. This gets complicated by the fact that the
metric of boson stars is not analytically known. However,
preliminary studies in the case of the simplest -black body
behaving- accretion disk, have shown that the spectrum of the
radiation emitted is modified, specially at high energies.

It has been already noted that {\it X-ray astronomy} can probe
regions very close to the Schwarzschild radius. Recent results
from the Japanese-US ASCA mission have revealed broadened iron
lines, a feature that comes from regions which are under strong
gravitational influence.  X-ray astronomy could be used to map out
in detail the form of the potential well, and then as a
discriminator. The NASA Constellation-X mission, to be launched in
2008, is optimized to study the iron K line feature discovered by
ASCA.

{\it Observations of very large baseline interferometry (VLBI)}
could also give the signature to discriminate among these models.
Due to the phenomenon known as ``shadowing'', we might expect some
diminishing of the intensity right in the center, this would be
provided by effects upon relativistic orbits, however, this will
not be as pronounced as if a black hole is present: for that case,
many photons are really gone through the horizon and this deficit
also shows up in the middle. Instead, if a boson star is there,
some photons will traverse it radially, and the center region will
not be as dark as in the black hole case. We also mention that the
project ARISE (Advanced Radio Interferometry between Space and
Earth) is going to use the technique of Space VLBI. It will study
gravitational lenses at resolutions of tens of $\mu$arcsecs,
yielding information on the possible existence (and signatures) of
compact objects with masses between $10^3$ and $10^6 M_{\odot}$.

If a particle with stellar mass is observed to spiral into a
spinning object with a much larger mass and a radius comparable to
its Schwarzschild length, from the emitted {\it gravitational
waves}, one could obtain the lowest multipole moments. The black
hole no-hair (or two-hair) theorem establishes that all moments
are determined by its lowest two, the mass and angular momentum
(assuming the charge equal to zero), for instance the mass
quadrupole moment would give $M_2=-L^2/M$. Should this not be so,
the central object would not be a black hole, and as far as we
know, the only remaining viable candidate would be a boson star.
In this case, all multipole moments are determined by the boson
star lowest three.

The formation of boson stars and black holes can be competitive
processes. Then, it might well be that even if we discover that a
black hole is in the center of the Galaxy, other galaxies could
harbor non-baryonic centers. In the case of boson stars, only
after the discovery of the boson mass spectrum we shall be able to
determine a priori which galaxies could be modeled by such a
center, if any. Observations of galactic centers could then
suggest the existence of boson scalars much before than their
discovery in particle physicists labs.

\section*{Acknowledgments}
D.F.T. was supported by CONICET as well as by funds granted by
Fundaci\'{o}n Antorchas. S.C. and G.L. were supported by MURST
fund (40\%) and art. 65 D.P.R. 382/80 (60\%). G.L. further thanks
UE (P.O.M. 1994/1999). D.F.T. further thanks E. Mielke for his
invitation to deliver this talk.


\begin{thebibliography}{99}

\bibitem{T}D. F. Torres, S. Capozziello, and G. Lambiase,
\Journal{\PRD}{62}{104012}{2000}.

\end{thebibliography}
\end{document}